\documentclass[12pt]{iopart}
\usepackage{graphicx}
\newcommand{\beq}{\begin{equation}}
\newcommand{\eneq}{\end{equation}}
\newcommand{\bea}{\begin{eqnarray}}
\newcommand{\enea}{\end{eqnarray}}

\begin{document}

\title{Frustration of decoherence in $Y$-shaped superconducting Josephson
networks}

\author{Domenico Giuliano$^{1}$ and Pasquale Sodano$^{2}$}

\address{$^1$ Dipartimento di Fisica, Universit\`a della Calabria and
             I.N.F.N., Gruppo collegato di Cosenza, Arcavacata di Rende
             I-87036, Cosenza, Italy
\\ $^2$ Max-Planck Institut f\"ur Physik Komplexer
Systeme, N\"othnitzer Strasse 38, 01167, Dresden, Germany \footnote{ 
{\it Permanent address} : Dipartimento di
Fisica, Universit\`{a} di
Perugia, and I.N.F.N., Sezione di Perugia, Via A. Pascoli, 06123,
Perugia, Italy}}

\ead{$^1$giuliano@fis.unical.it \; , \; $^2$sodano@pg.infn.it}
\begin{abstract}
We examine the possibility that pertinent impurities in a
condensed matter system may help in designing quantum devices with enhanced
coherent behaviors. For this purpose, we analyze a field theory model
describing Y- shaped superconducting Josephson networks. We show that   a new
finite coupling stable infrared fixed point emerges in its phase diagram; we
then explicitly  evidence that, when engineered to operate near by this new
fixed point, Y-shaped networks support two-level quantum systems, for which the
entanglement with the environment is frustrated. We briefly address the
potential relevance of this result for engineering finite-size 
 superconducting devices with enhanced quantum coherence. 
  Our approach uses boundary conformal field theory since it  naturally allows 
for a field-theoretical treatment of the phase slips (instantons), 
 describing the quantum tunneling between  degenerate levels.

\end{abstract}

\pacs{71.10.Hf, 74.81.Fa,  11.25.hf, 85.25.Cp}
\maketitle

 For engineering quantum devices one has often to tame the decoherence arising
from the interaction of a pertinent two-level system with both the control
circuitry and the quantum modes lying outside the subspace spanned by the two
operating states.  An important  source of decoherence  arises when  
the total state of the two-level system and of its environment evolve 
towards an entangled state. 
If a system is coupled to more than one bath, and its
entanglement with each one of the baths is suppressed by the
other(s),  decoherence may be  frustrated \cite{novais,kohler}. 
In this paper, we evidence how frustration of
decoherence  may arise from the existence of a finite coupling fixed
point (FFP) in the phase diagram of the quantum theory describing
the device. 

Existence of finite coupling fixed points in  condensed matter
is a rare instance realized, so far, only in quantum systems with 
pertinent impurities. Remarkable examples of systems exhibiting 
attractive FFP's are provided by the two-channel single-impurity \cite{blandin}
and two-impurity \cite{jones} overscreened Kondo models, as well as by
$Y$-shaped junction of quantum wires \cite{aoc}. At variance, $Y$-shaped
junctions of one-dimensional atomic condenstates \cite{Demler} exhibit a 
repulsive FFP, signaling the existence of a new  transition point
 between stable  weakly and the strongly coupled phases. 

\begin{figure}
\includegraphics*[width=0.60\linewidth]{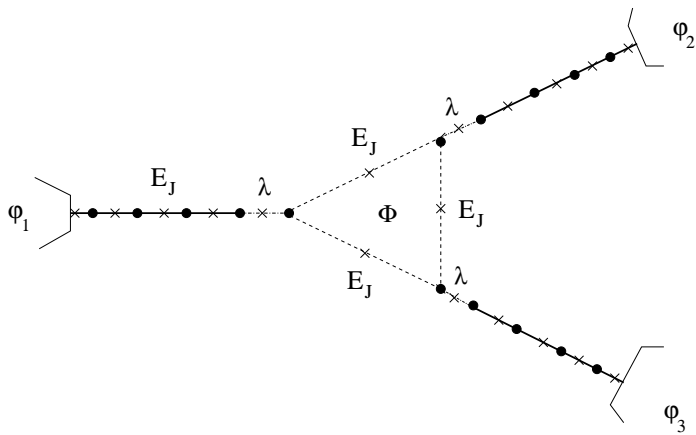}
\caption{Sketch of the YJJN; {\bf Inset}:
graphical exact solutions for the energy levels at $g=9/8$.}
\label{fig1}
\end{figure}
\normalsize

Boundary conformal field theories \cite{cardylavoro} are a natural 
setting to investigate stable phases and phase transitions of
quantum impurity systems, once the quantum impurity  is traded 
\cite{cardylavoro} for a boundary interaction, involving only a
subset of the {\it bulk} degrees of freedom: the boundary interaction
is then renormalized by the bulk degrees of freedom, and the infrared (IR) 
behavior is determined by the stable fixed point(s) in the phase diagram.

Superconducting Josephson devices are not only  promising
candidates for realizing quantum coherent two-level systems \cite{shon0},
but also provide remarkable realizations of quantum systems with 
impurities, whose phase diagrams, in the simple cases so far investigated, 
admit only two fixed points: an unstable weak
coupling fixed point (WFP), and a stable one at strong coupling
(SFP) \cite{giuso1}. The approach developed in Ref.\cite{giuso1}
naturally allows for a field-theoretical treatment of the phase slips 
describing quantum tunneling between degenerate levels, and provides 
remarkable analogies to models of quantum Brownian motion on 
frustrated planar lattices \cite{kaneyi,saleurb}. 
When an effective two-level quantum system is
operated near by the WFP or the SFP, there is no frustration of
decoherence, since, at strong coupling, there is not
even quantum tunneling between the degenerate states while, at weak
coupling, there is full entanglement between the two degenerate
states and the plasmon modes. In the following,  
we shall show that a FFP emerges in a $Y$-shaped Josephson junction 
network (YJJN), and that it may be pertinently used to engineer two-level
systems  with enhanced quantum coherence. 

A YJJN is realized by joining a circular Josephson junction array 
{\bf C} to three finite Josephson chains via weak links of nominal strength 
$\lambda$ (see Fig.\ref{fig1}).  {\bf C} is pierced 
by a dimensionless magnetic flux $\Phi$, and is joined to one of the endpoints
of the three chains (inner boundary); the other endpoints (outer boundary)
are connected to three bulk superconductors at fixed phases  
$\varphi_i$ ($i=1,2,3$). For simplicity, we assume that all the junctions in
the YJJN are of strength $E_J$ and that $\lambda \ll E_J$. The 
Hamiltonian describing {\bf C} is given by

\beq 
H_{\bf C} =
\frac{E_c}{2} \sum_{ i = 1}^3 \left[ - i \frac{
\partial }{
\partial \phi_0^{(i)} } - W_g \right]^2 - 2 E_J \sum_{ i = 1}^3 \cos \left[
\Delta \phi_0^{(i)}   + \frac{ \Phi }{3} \right] 
\;\;\;\;, \label{eisl} 
\eneq 
\noindent 
where $\Delta \phi_0^{(i)}=
\phi_0^{(i)}-\phi_0^{(i+1)}$, $\phi_0^{(i)}$ is the phase of the
superconducting order parameter at grain $i$, 
and $W_g$ is a gate voltage. If $E_J / E_c \ll 1$, $ W_g^{(i)}  =
N   + \frac{1}{2} + h$, with integer $N$ and $0<h < 1/2$, the
low-energy dynamics is governed only by the two states with total
charge equal to $N$ and to $N + 1$.

The procedure outlined in Ref.\cite{giuso1} allows 
to describe the three finite chains with a Tomonaga-Luttinger 
Hamiltonian 

\beq
H_0 = \frac{g}{4 \pi} \sum_{i = 1}^3 \: \int_0^L \: d x \: \left[
\frac{1}{v} \left( \frac{ \partial \Phi_i}{ \partial t} \right)^2
+  v \left( \frac{ \partial \Phi_i}{
\partial x} \right)^2 \right] 
\:\:\:\: .
\label{plet2} 
\eneq
\noindent 
In Eq.(\ref{plet2}) $\Phi_i$ describe the plasmon modes of the chains, 
and $g$ and $v$ depend on the constructive parameters of the network
\cite{giuso1}.

Fixing the phase at the outer boundary of the
chains sets   Dirichlet boundary conditions on $\Phi_i (x )$ at
$x=L$: $\Phi_i ( L ) = \varphi_i$. Since we require that 
the charge tunneling between ${\bf C}$ and the inner
boundary of the three chains is described by  a Josephson-like
interaction, with nominal strength $\lambda \ll E_J$, one should use
Neumann boundary conditions at the inner boundary, i.e.
$\frac{\partial \Phi_i ( 0 )}{\partial x} = 0$ $\forall i$. This
allows to  write the tunneling Hamiltonian as $H_T = - \lambda
\sum_{ i = 1}^3 \cos [ \Phi_{i} (0) - \phi_0^{(i)} ] $. 

A boundary field theory approach allows to trade 
$H_{\bf C} + H_T$ with an
effective boundary Hamiltonian, $H_b$, involving only $\Phi_i
(0)$, and given by 

\beq 
H_b = - 2 \bar{E}_W \sum_{ i = 1}^3 : \cos
[ \vec{\alpha}_i \cdot \vec{\chi} ( 0 )  + \gamma ]: 
\;\;\;\; ,
\label{let3} 
\eneq 
\noindent
with $\chi_1 (x)=  \frac{1}{\sqrt{2}}
[\Phi_1 (x ) - \Phi_2 (x)]$, $\chi_2 (x)=  \frac{1}{\sqrt{6}}
[\Phi_1 (x ) + \Phi_2 (x) - 2 \Phi_3 (x) ]$,   $\vec{\alpha}_1 = (
1 , 0 )$, $\vec{\alpha}_2 = ( - \frac{1}{2} , \frac{ \sqrt{3}}{2}
)$, $\vec{\alpha}_3 = ( - \frac{1}{2} , - \frac{ \sqrt{3}}{2} )$,
$\gamma = \tan^{-1}  [ 3 \tan ( \frac{ \Phi }{3}  )]$, and
$\bar{E}_W = \left( \frac{a}{L} \right)^\frac{1}{g} E_W $, with
$E_W \approx \frac{ \lambda^2 E_J}{ 24 (E_c)^2 \tilde{h}^2} \sqrt{
1 + 2 \sin^2 ( \frac{ \Phi }{3} ) } $. The colons $: \: :$
denote normal ordering with respect to the ground state of the
plasmon modes, $ | \{ 0 \} \rangle$. In the following, we shall argue 
that, for $\gamma = \pi / 3$, there is a finite range of values of 
$g$, for which a YJJN supports a FFP: this results from the fact
that, for this value of $\gamma$,  the two plasmon baths 
$\chi_1$ and $\chi_2$, cooperate to destabilize  both the SFP and the WFP. 

\begin{figure}
\includegraphics*[width=1.0\linewidth]{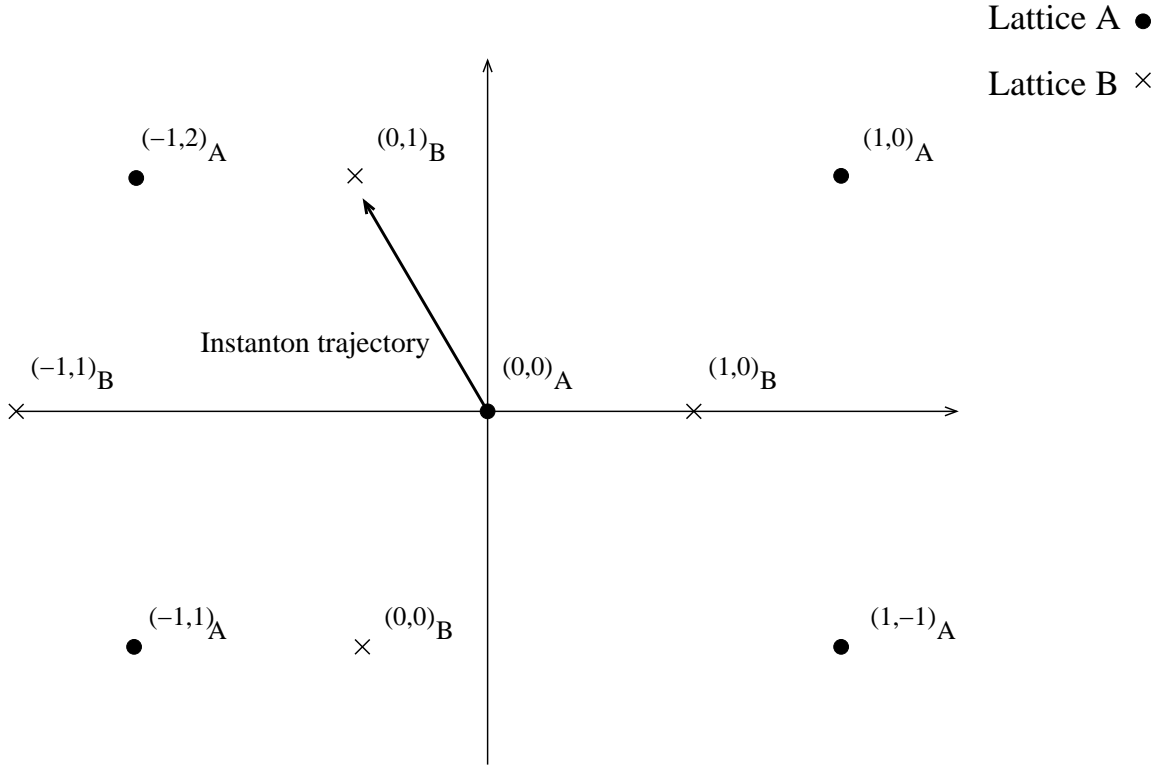}
\caption{Some points on the two lattices A and B: at
$\gamma = \pi / 3$, the two lattices are degenerate. In
this case, the set of the minima of the boundary
potential spans a honeycomb lattice, whose sites are
connected  by instanton trajectories as the one drawn in the figure.}
\label{ablat}
\end{figure}

The perturbative second-order renormalization group 
(RG) equation for the running coupling strength $G = L \bar{E}_W$, 
given by 

\beq
\frac{ d G }{ d \ln ( \frac{L}{L_0} )} = \left( 1 -
\frac{1}{g} \right) G - 2 G^2 
\:\:\:\: , 
\label{additional1}
\eneq
\noindent
shows that $H_b$ is  a relevant perturbation for
$g>1$, while it is irrelevant for $g<1$. In Eq.(\ref{additional1}), 
$L_0$ is a pertinent reference length scale.
The strongly coupled fixed point (SFP) is reached when the running 
coupling constant  $G$ goes to $\infty$. The fields $\chi_j ( x )$, $j=1,2$,
now obey Dirichlet boundary conditions at $x=0$ and 
$\chi_1 ( 0 ) , \chi_2 ( 0 )$ are determined by  the manifold of
the minima of the effective boundary potential  (Eq.(\ref{let3})). 
One sees that for  $0\leq \gamma < \pi / 3$, the minima lie on
the triangular sublattice A, defined by $ (\chi_1 (0)  , \chi_2  ( 0 ) ) =
( 2 \pi m_{12}  , \frac{2}{ \sqrt{3}} [ 2 \pi m_{13}  + \pi m_{12} ] ) $, 
while, for $ \pi / 3 < \gamma \leq 2 \pi / 3 $,
the minima lie on the triangular sublattice B, given by
$ (\chi_1 (0), \chi_2 (0) ) = ( 2 \pi m_{12} - \frac{2 \pi}{3} ,
\frac{2}{ \sqrt{3}} [ 2 \pi m_{13}  + \pi m_{12} - \pi ] )$, with $m_{12},
m_{13}$ relative integers. From Eq.(\ref{let3}), one sees also that the 
 difference in energy between the sets of the minima forming the A and B
sublattices is given by $\sim \bar{E}_J \sin ( \gamma - \frac{\pi}{3} )$.
The manifold of the minima is depicted in Fig.\ref{ablat}, where the
instanton connecting the  degenerate minima of the honeycomb lattice emerging
when the A and B sublattices are degenerate, is shown.

Following the approach outlined in Ref.\cite{kaneyi}, instanton effects 
may be taken into account through 

\beq
\tilde{H}_b = - Y \sum_{ i =1}^3 \{ T^- V_i ( 0 ) + T^+ \bar{V}_i
( 0 ) \}
\:\:\:\: .
\label{additional2}
\eneq
\noindent
In Eq.(\ref{additional2}), $\vec{T}$ is an effective isospin operator, 
connecting two neighboring minima of the honeycomb lattice of 
the zero-mode eigenvalues, 
 $V_i (\bar{V}_i ) = : \exp \left[ - (+) \frac{2}{3} \vec{\alpha}_i \cdot 
\vec{\Theta} \right]:$, with $\vec{\Theta}$ being the dual fields of
$\vec{\chi}$, while  $\tilde{H}_b$ is  the "dual" boundary Hamiltonian of
Eq.(\ref{let3}). $Y$ is an effective coupling  defined as $Y = E_J - \bar{E}_W$
\cite{tsvelik}. From the O.P.E. of the vertex operators entering $\tilde{H}_b$,
the RG equation for the running coupling strength $ y = L Y$ is

\beq
\frac{ d y}{ d \ln \left( \frac{L}{L_0} \right) }
= \left( 1 - \frac{4g}{9} \right) y - \frac{2g}{3} y^3
\:\:\:\: . 
\label{additional4}
\eneq
\noindent
For $\gamma = \pi / 3 $ and $1 < g < \frac{9}{4}$, neither
the WFP, or the SFP, are stable. Accordingly, a minimal hypothesis
for the phase diagram requires a FFP at $y = y_*$, with $y_*$
finite. For instance, for $g = \frac{9}{4} - \epsilon$,
 with $\epsilon \ll 1$, one obtains $y^* \approx \left( \frac{2}{3}
\right)^\frac{1}{2} \sqrt{\epsilon}$. 

The energy of the minima may be varied by changing the phases of 
the three bulk superconductors since the eigenvalues $(p_1 , p_2 )$ 
of the zero-modes of the fields $\chi_j $ (it obeys Dirichlet b.c.!) 
depend on the phases $\varphi_j$ as  

\beq
(p_1 , p_2 ) = \sqrt{ \frac{g}{2}} ( n_1 + \sqrt{2} \beta_1 ,
\frac{1}{\sqrt{3}} ( 2 n_2 + n_1 + 2 \sqrt{2} \beta_2 ) )
\:\:\:\: , 
\label{sla}
\eneq
\noindent
on sublattice A, and 

\beq
 (p_1 , p_2 ) = \sqrt{ \frac{g}{2}} ( n_1 - \frac{1}{3} +
\sqrt{2} \beta_1 , \frac{1}{\sqrt{3}} ( 2 n_2 + n_1 -1 + 2
\sqrt{2} \beta_2 ) )
\:\:\:\: ,
\label{slb}
\eneq
\noindent
on sublattice B. In Eqs.(\ref{sla},\ref{slb}) $ (n_1 , n_2 )$ are integers,
  $\beta_1 = ( \varphi_1 - \varphi_2 ) / ( 2 \pi \sqrt{2})$, and
$\beta_2 = ( \varphi_1 + \varphi_2 - 2 \varphi_3 ) / ( 2 \pi
\sqrt{6})$.

As it happens with other superconducting systems \cite{shon0}, 
also a YJJN supports a two-level quantum system, operating between
two pertinently selected quantum states. Indeed, for
$y \ll 1$ and near the SFP, the low-energy spectrum is given by $E
= \frac{ \pi v }{ 2 L } [ \vec{p} ]^2 + E'$, where $\vec{p} = (
p_1 , p_2 )$ labels the zero-mode contribution, while $E'$ comes
from the  plasmon modes: thus, for $\gamma = \pi / 3$, a pertinent
tuning of $\beta_1$ and $\beta_2$ renders degenerate the zero-mode
contributions to the total energy coming from two nearest neighboring sites
of the honeycomb lattice resulting from the degeneracy of the A and B
sublattices (see Fig.\ref{ablat}). This  happens, for instance, if
$\beta_1 = 1 / 3 \sqrt{2} , \beta_2 = 0$: the two degenerate
quantum states $|\uparrow \rangle$ and $| \downarrow \rangle$
-labelled by $( n_1 , n_2 ) = (0,0)$ on sublattice A and by $ (n_1
, n_2 ) = ( 1 , 0 )$ on sublattice B- are macroscopically 
characterized by the opposite values of the Josephson current flowing
across chain-1 and chain-2, namely:  $I_1 = - I_2 = \pm \frac{ \pi
g v e^*}{3 L}$, $I_3 = 0 $. 

Quantum tunneling between the degenerate states is  induced by
$\tilde{H}_b$, with matrix element $-Y$. Setting $\beta_2 = 0$,
and $\beta_1 = 1 / 3 \sqrt{2} + \delta/ ( 2 \pi)$, with $\delta /
2 \pi \ll 1$, one easily gets an effective Hamiltonian for the
two-level quantum system as

\beq 
H_2 = \epsilon_0 ( \delta ) {\bf
I} + \epsilon ( \delta ) \sigma^z - Y \sigma^+ \bar{V}_1 ( 0 ) 
-   Y \sigma^- V_1 ( 0 ) \;\;\;\; .
\label{eqadi1} 
\eneq 
\noindent
In Eq.(\ref{eqadi1}) $\epsilon_0 ( \delta ) = \frac{g}{2} \left( \frac{1}{9}
+ \frac{\delta^2}{ 4 \pi^2 } \right)$, $\epsilon ( \delta ) =
\frac{g}{3} \frac{\delta}{ \sqrt{2 } \pi } $, the $\sigma^a$'s are
the Pauli matrices,  $\delta$ is a control parameter determined
by the phases $\{\varphi_i \}$, and $- Y [ \sigma^+ \bar{V}_1 ( 0 ) + 
\sigma^- V_1 ( 0 ) ]$   describes
the interaction of the two-level system with the phase slip operators 
introduced in Eq.(\ref{additional2}) 

In the spin Hamiltonian describing the two level system in Eq.(\ref{eqadi1}),
one sees that there is a $z$-component proportional to $\epsilon ( \delta )$, 
as well as an $x$-component proportional to $Y$. While $\epsilon ( \delta )$
does not get renormalized by the interaction with the two plasmon fields,
$Y$ is renormalized  and its value measures the amount of entanglement 
between the two-level system and the plasmon modes bath. In particular,  
if $Y$ is irrelevant,  the two-level system decouples from the environment
and behaves as a classical (Ising-like) spin, pointing along $z$. 
When this happens, no energy is dissipated into the environment, 
and the spectrum of the Hamiltonian in  Eq.(\ref{eqadi1}) is given by
two classical states with $\omega = \pm \epsilon ( \delta )$. 
If $Y \to \infty$, the effective field acting on the 
two-level system would again make it behave as a classical spin, pointing
in the $x$-direction: now, all the energy is dissipated into the
environment and the spectrum of  Eq.(\ref{eqadi1}) is given by only
an overdamped mode at $\omega = 0$. Only when $Y$ takes a finite 
value $y_*$, the competition between $\epsilon ( \delta )$ and $y_*$ may 
lead to the emergenge of the frustration of the decoherence of the two-level
system, since  there is the possibility that, for a pertinent choice of
the control parameter $\epsilon ( \delta )$, there is a finite damping,
resulting in two broad modes, centered at pertinent renormalized energies.
 
\begin{figure}
\includegraphics*[width=1.0\linewidth]{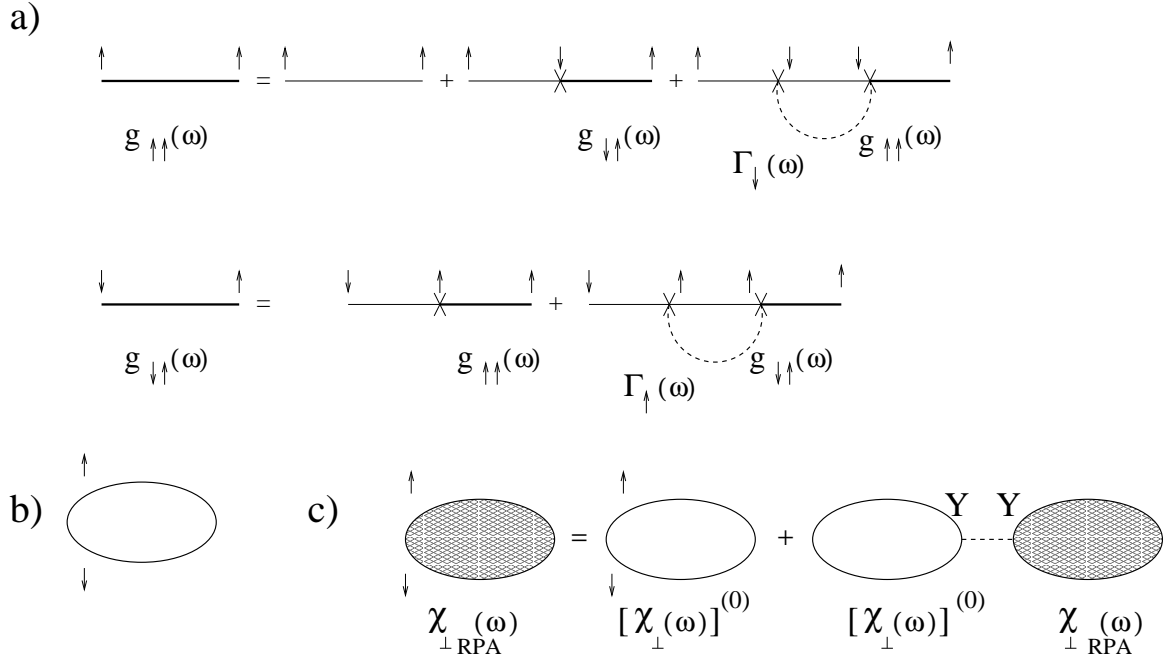}
\caption{{\bf a):} Graphical representation for the Schwinger-Dyson 
equations for $g_{\sigma \sigma'} ( \omega)$; {\bf b):} The ``bubble''
yielding the dynamical spin susceptibility; {\bf c):} Graphical representation
of the RPA summation implemented to compute $\chi_\perp ( \omega )$ near
the FFP.}
\label{dys1}
\end{figure}

To evidence the frustration of decoherence around the FFP, 
we compute the spectral density of  the Hamiltonian in  
Eq.(\ref{eqadi1}), given by $\chi^{``}_\perp 
(  \omega )  / \omega$, where  $\chi^{``}_\perp (  \omega )$ 
is the imaginary part of the transverse
dynamical spin susceptibility \cite{novais}. 
The diagrams contributing to  $\chi_\perp ( \omega )$ are  
shown in Fig.(\ref{dys1} {\bf b}):  $\chi_\perp ( \omega )$ is computed as 
a loop defined by the $| \uparrow \rangle$-state propagating forward in 
(imaginary) time, and by  the $| \downarrow \rangle $-state propagating 
backward. It is given by

\beq
\chi_\perp ( \omega ) = - i \int_{ - \infty}^\infty \; \frac{dz}{2 \pi } \{ 
g_{\uparrow , \uparrow}^* ( - z ) g_{\downarrow , \downarrow} ( z + \omega )  +
 g_{\downarrow , \downarrow}^* ( - z )   g_{\uparrow , \uparrow} ( z + \omega
 )  \}
\:\:\:\: ,
\label{i7}
\eneq
\noindent
where $g_{ \sigma \sigma} ( \omega )$ is the Fourier tranform of the 
 propagator of the ``spin'' eigenstate  $ | \sigma \rangle$ 
($\sigma = \uparrow , \downarrow$). 

For  $\gamma = \pi / 3$ and 
$g> 9/4$, the boundary interaction is irrelevant, and one may  
neglect corrections to the
amplitudes of order $Y^2$. This amounts  to substituting 
$g_{ \sigma \sigma} ( \omega )$ with  its noninteracting limit,
 $ g^{(0)}_\sigma  ( \omega ) = 
1/ [ i \omega + \sigma \epsilon ( \delta ) ]$, yielding

\beq
\frac{ \chi^{"} ( \omega )}{\omega } 
\propto [ \delta ( \omega + 2 \epsilon ( \delta )  ) -
\delta (\omega - 2  \epsilon ( \delta ) ) ] / \omega
\:\:\:\: .
\label{addi.2.1}
\eneq
\noindent
Eq.(\ref{addi.2.1}) shows that there is no entanglement (for $g> 9 / 4$) 
between the two-level quantum
system and the plasmon modes. Since, in this range of $g$, there
is no tunnel splitting between the two degenerate states, the system
is classical and no quantum coherence emerges.

For $\gamma = \pi / 3$ and $g <1$,  instantons provide a 
relevant perturbation and  the IR behavior of the system is driven by the  
WFP. To compute  $\chi_\perp ( \omega )$, one now needs to substitute
$ g^{(0)}_\sigma  ( \omega ) $ with the dressed propagator,
$g_{ \sigma \sigma} ( \omega )$, drawn in  Fig.(\ref{dys1} {\bf a}), where
the solid heavy line represents the fully dressed propagator, 
while the solid light line represents $g_\sigma^{(0)} ( \omega )$.  
As a result,  $g_{ \sigma \sigma'} ( \omega )$ is given by

\beq
g_{\sigma , \sigma'} ( \omega ) = \frac{ \delta_{\sigma , \sigma'}
[ [ g^{(0)}_{\bar{\sigma}} ]^{-1} ( \omega ) + Y^2 \Gamma_\sigma (
\omega ) ] + i Y \delta_{ \sigma , {\bar{\sigma}}{'} }  }{ \{ 
[g^{(0)}_\uparrow ]^{-1} ( \omega ) + Y^2 \Gamma_\downarrow ( \omega ) \}
 \{ [ g^{(0)}_\downarrow ]^{-1} ( \omega ) + Y^2 \Gamma_\uparrow ( \omega ) \}
+ Y^2  } 
\:\:\:\: , 
\label{eqle3} 
\eneq 
\noindent 
with $\bar{\sigma }  =  - \sigma$, and  $ \Gamma_\sigma ( \omega )$ is  the 
Fourier transform of the propagation function

\beq
 \Gamma ( \tau_1 - \tau_2 ) = \langle \{ 0
\} | : e^{ \pm i \frac{2}{3} \Theta ( \tau_1 )}: : e^{ \mp i
\frac{2}{3} \Theta ( \tau_1 )}:  | \{ 0 \}  \rangle = [ e^{ \frac{
\pi i }{L} v \tau_1 } - e^{ \frac{ \pi i }{L} v (\tau_2 + i \eta)
}]^{ - \frac{8}{9} g }
\:\:\:\: , 
\label{i6}
\eneq
\noindent
at frequency $\omega - \sigma \epsilon ( \delta ) $. 
$\chi_\perp ( \omega )$ near the WFP is computed by taking the 
large-$Y$ limit of Eq.(\ref{eqle3}), yielding 

\beq
\frac{\chi^{"}  ( \omega )}{ \omega } \propto [ | 2 \epsilon
(\delta ) + \omega |^{3 - \frac{16}{9} g } - | 2 \epsilon (\delta
) - \omega |^{3 - \frac{16}{9} g}]/\omega 
\:\:\:\: . 
\label{ii6}
\eneq
\noindent
Eq.(\ref{ii6}) shows that  a large part of the spectral weight has 
moved now from the side peaks towards $\omega = 0$, signaling the  strong
decoherence of the two-level system described by Eq.(\ref{eqadi1}). 

\begin{figure}
\includegraphics*[width=0.60\linewidth]{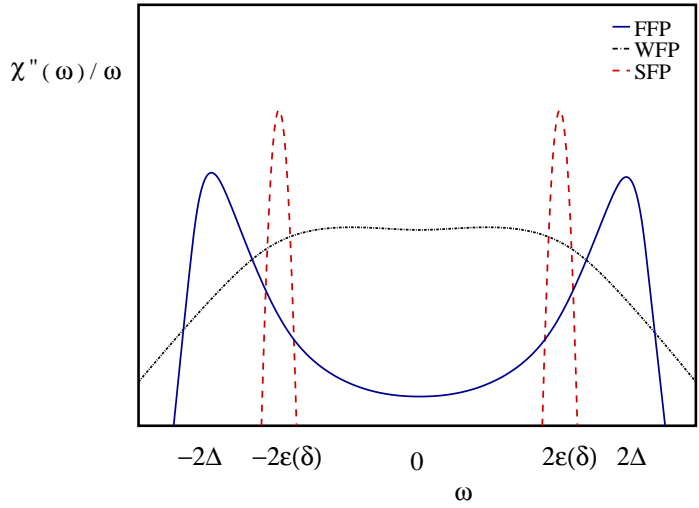}
\caption{Qualitative behavior of $\chi^{"} ( \omega ) / \omega$ in the
various regimes.}
\label{fig2}
\end{figure}

For $\gamma = \pi / 3$ and 
$1 < g < \frac{9}{4}$, the  IR behavior is driven by the FFP. A 
closed-form computation of $\chi_\perp ( \omega )$ is now possible 
only for special values of $g$.  For  instance, if 
$g = \frac{9}{4} - \epsilon$
with $\epsilon  \ll 1$,  $y_*$ is $ \ll 1$, and  one may compute  $\chi_\perp 
( \omega )$  by  resorting to a RPA summation, graphically sketched in
Fig.(\ref{dys1}{\bf c)}). The result is

\[
[ \chi_\perp ]_{\rm RPA}  ( \omega ) \approx \frac{1}{ \omega - 2 \sqrt{ 
\epsilon^2 ( \delta ) + Y^2} - Y^2 \Gamma [ - 1 - \frac{8}{9}  \gamma ] 
( - \omega )^{ 1 + \frac{8}{9}  \gamma } }
\]
\beq
 + \frac{1}{ \omega + 2 \sqrt{ 
\epsilon^2 ( \delta ) + Y^2} - Y^2 \Gamma [ - 1 - \frac{8}{9}  \gamma ] 
( - \omega )^{ 1 + \frac{8}{9}  \gamma } } 
\:\:\:\: . 
\label{ii1}
\eneq
\noindent
When writing  $[ \chi_\perp ]_{\rm RPA}  ( \omega )$ as a function of the
dimensionless variable $x = \frac{2L}{ \pi v } \omega$, taking into 
account that the dimensionless variable $y = Y L^{ 1 - \frac{4}{9}g} 
\longrightarrow y_*$ as $L \to \infty$, one gets

\beq
[ \chi_\perp ]_{\rm RPA}  ( x ) \propto \frac{ 1}{ x - 2 \Delta + e^{ i \pi 
\frac{8}{9}  \gamma } x^{  1 + \frac{8}{9}  \gamma } } +
\frac{ 1}{ x + 2 \Delta + e^{ i \pi 
\frac{8}{9}  \gamma } x^{  1 + \frac{8}{9}  \gamma } } 
\:\:\:\: , 
\label{ii4}
\eneq
\noindent
where $\Delta = \frac{ 2 L }{ \pi v } \sqrt{ E^2 + Y^2 }$. The 
imaginary part of Eq.(\ref{ii4}) shows two peaks centered
around $\pm 2 \sqrt{ [ \epsilon ( \delta ) ]^2 + (\frac{ \pi v}{L}
y_*)^2 }$, where $y_*$ is the finite fixed point value of the running
coupling constant. In Fig.\ref{fig2}, we provide the plot 
of $\chi^{``} ( \omega ) / \omega$. 

A very special situation arises for $\gamma = \pi /3$ when $g = 9 / 8$
since, for this value of $g$, the scaling dimension of the relevant instanton 
operators equals 1/2, just as it happens with  fermionic operators.  
Indeed, for $g = 9 / 8$, the plasmon field $: \exp \left[ - \frac{2}{3} i 
\sqrt{2g} \Theta ( x , \tau ) \right] :$ becomes a fermionic operator
$ \psi ( x + i v \tau ) $ ($- L \leq x \leq L$), and
the spin-1/2 operators may be fermionized according to

\beq
\sigma^z \longrightarrow a^\dagger a - 
\frac{1}{2} \;\;\; , \;\;  \sigma^+ \longrightarrow 
a^\dagger  e^{ \frac{3}{2} i \pi \sqrt{ \frac{2}{g}} P_1 } 
\;\;\;\; , 
\label{ffer3}
\eneq
\noindent
where the zero-mode operator $P_1$ ensures, for $g = 9 / 8$, the
correct anticommutation relations between $\psi $ and the operators
in Eq.(\ref{ffer3}). As a result, the  two-level Hamiltonian  
Eq.(\ref{eqadi1}) becomes

\beq
H_{\rm Fer} = - i v \int_{-L}^L \: d x \: \psi^\dagger (x ) \frac{ \partial
\psi ( x )}{ \partial x} + \epsilon ( \delta ) ( a^\dagger a - 
\frac{1}{2} )  - Y \psi ( 0 )
a^\dagger  e^{ \frac{3}{2} i \pi \sqrt{ \frac{2}{g}} P_1 }   -  {\rm h.c.}
\:\:\:\: ,
\label{ffer1b}
\eneq
\noindent
with twisted boundary condition 
\beq
\psi ( L ) = \exp \left[ \frac{4}{3} i \pi \sqrt{g} p_1 \right] \: \psi (-L)
\:\:\:\: , 
\label{ffer2}
\eneq
\noindent
where $p_1$ is the eigenvalue of the zero-mode operator $P_1$. 
A similar situation arises in the analysis of a spin-1/2 Kondo system at the 
Toulouse point \cite{emery}. In particular, the Hamiltonian in 
Eq.(\ref{ffer1b}) has been recently proposed \cite{degiovanni} to describe 
two qubits  at the end of a finite length 1d cavity.

To determine the energy eigenstates of the Hamiltonian in Eq.(\ref{ffer1b}),
$ | E \rangle$, with the boundary condition in Eq.(\ref{ffer2}), one may
posit

\beq
| E \rangle = \left\{ \int_{-L}^L \: d x \: f_E ( x ) \psi^\dagger ( x ) 
+ \lambda_E a^\dagger \right\} | {\bf 0} \rangle 
\:\:\:\: ,
\label{ffer4}
\eneq
\noindent
where  $ | {\bf 0} \rangle$ is  the simultaneous
eigenstate of $P_1$ and $\sigma^z$ given by $| ( p_1 = 
- \sqrt{ \frac{g}{2}} \frac{ \delta_1 }{ 2 \pi} ) ,  \downarrow \rangle$. 
From  $ H_{\rm Fer} | E \rangle = E | E \rangle$, one gets 

\begin{eqnarray} 
- i v \frac{ \partial f_E (x )}{ \partial x} + Y \lambda_E e^{ \frac{3}{4} i
\delta_1} \delta ( x ) = E f_E ( x ) \nonumber \\
\lambda_E \epsilon ( \delta ) + Y f_E ( 0 ) e^{ - \frac{3}{4} i
\delta_1} = E \lambda_E 
\:\:\:\: , 
\label{ffer5}
\end{eqnarray}
\noindent
which is solved by

\beq
f_E ( x ) = \frac{1}{ \sqrt{2L}} [ e^{ i \frac{ \chi}{v} } \theta ( x ) +
 e^{ - i \frac{ \chi}{v} } \theta ( - x ) ] 
\:\:\:\: , 
\label{ffer6}
\eneq
\noindent
provided that 

\beq
\frac{E}{v} L + \frac{ \chi}{v} + \frac{\pi}{2} = 0 
\:\:\: ; \;\; v \tan \left( \frac{ \chi}{v} \right) + L 
\frac{ Y^2 }{ E - \epsilon ( \delta )}
= 0 \;\;\;\; . 
\label{ffer7}
\eneq
\noindent
In  the inset of Fig.\ref{fig1}, Eqs.(\ref{ffer7}) are graphically
solved for  $\epsilon ( \delta ) = 0$, using the  
dimensionless variable $x = LE / v$. 

For $g>1$, inhomogeneities in the fabrication parameter $E_J$
 provide an irrelevant perturbation, since the pertinent
operator scales as $ \left( \frac{L}{L_0} \right)^{1-g}$
\cite{giuso1} and,  thus, does not alter the main 
results of our analysis. Furthermore, today 's technology allows to fabricate 
superconducting devices with values of $g$ ranging from 
$g < 1$, to $g \sim 2$ \cite{haviland}.

\begin{figure}
\includegraphics*[width=0.60\linewidth]{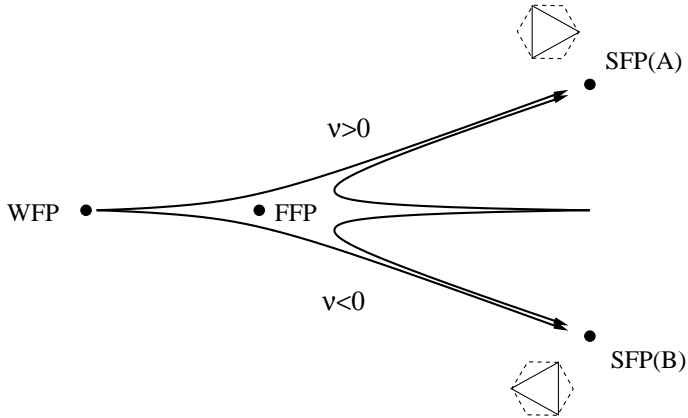}
\caption{Sketch of the RG flow for $\gamma = \pi / 3 + \nu$ ($\nu / \pi
\ll 1$).} \label{figthree}
\end{figure}

Operating a YJJN near the FFP allows to 
engineer a realistic finite-size  two-level quantum device with 
enhanced quantum coherence. Indeed, 
for a YJJN of finite size $L$, the FFP is stable
against small fluctuations of the flux $\Phi$, provided
that $v /L$ is sufficiently big: if $\gamma = \pi / 3$ is
displaced by a small amount $\nu$, $v / L $ needs to be larger than
the energy splitting  $\bar{E}_W \sin ( \nu )$ between the minima of the two 
triangular sublattices. When $v/L < \bar{E}_W \sin (
\nu)$, there is a flow towards the SFP and,  depending on
${\rm sgn} (\nu)$, the minima of the boundary potential lie on
either one of the triangular A and B  sublattices (see  Fig.\ref{figthree}). 
The parameters $\beta_1 , \beta_2$ may be {\it safely} 
tuned to the degeneracy values
$\beta_1 = 1 / ( 3 \sqrt{2} )$, $\beta_2 = 0 $ by resorting to 
multipolar magnetic coils \cite{granata} inserted in loops connecting the
bulk superconductors at the outer boundary of the YJJN since,  for 
sufficiently long chains, the magnetic flux generated by the coil
does not alter the flux threading the circular Josephson junction 
array {\bf C}.

Josephson networks where $n$ finite chains are connected to a 
central circular array ${\bf C}$ share properties similar to a 
YJJN. For $n=4$, the resulting network is the tetrahedral qubit proposed
in Ref.\cite{blatter}.

In summary, our analysis of YJJNs provides an explicit example of 
a situation in which quantum impurities  may be pertinently used for 
engineering quantum devices with enhanced quantum coherence.

\noindent

\noindent 

\vspace*{0.2cm}

 We thank I. Affleck, C. Chamon, P. Degiovanni, R. Russo  and 
A. Trombettoni for  useful discussions and correspondence.

\end{document}